\def\be{\begin{equation}}
\def\te{\end{equation}}
\def\bea{\begin{eqnarray}}

\def\tea{\end{eqnarray}}

\def\ha{{1\over 2}}
\def\tt{\tilde t}
\def\tr{\tilde r}
\def\tS{\tilde S}

\documentstyle[12pt]{article}

\makeatletter                    
\@addtoreset{equation}{section}  
\makeatother                     

\begin{document}

\title{Hawking-Unruh Thermal Radiance as Relativistic Exponential
Scaling of Quantum Noise
\thanks{Invited talk at the Fourth International Workshop on Thermal Field
Theory and Applications, Dalian, China, August, 1995. Proceedings edited
by Y. X. Gui and F. C. Khanna (World Scientific, Singapore, 1996)}}
\author{B. L. Hu\\
{\small Department of Physics, University of Maryland,
College Park, MD 20742, USA }}
\date{\small {\it umdpp 96-52, gr-qc/yymmdd, Jan. 26, 1996}}
\maketitle
\begin{abstract}
The Hawking-Unruh effect of thermal radiance from a black hole or observed
by an accelerated
detector is usually viewed as a geometric effect related to the existence
of an event horizon. Here we propose a new viewpoint, that
the detection of thermal radiance in these systems
is a local, kinematic effect arising from the vacuum 
being subjected to a relativistic exponential scale transformation.
This kinematic effect alters  the relative weight of quantum
versus thermal fluctuations (noise) between the two vacua. This approach can
treat conditions which the geometric approach cannot, such as systems
which do not even have an event horizon. An example is the case of an observer
whose acceleration is nonuniform or only asymptotically uniform.
Since this approach is based on concepts and
techniques of non-equilibrium statistical mechanics, it is more adept to
dynamical problems, such as the dissipation, fluctuation, and
entropy aspects of particle creation and phase transitions in black hole
collapse and in the early universe.
\end{abstract}

\newpage

\section{Introduction}

The conventional view of the Hawking-Unruh effect \cite{Haw75,Unr76}
(thermal radiation observed by
an accelerated observer \cite{Unr76}, from a moving mirror \cite{FulDav},
or black hole \cite{Haw75,HawRad75} and for certain observers in an
inflationary universe \cite{GibHaw,Par76,Lap78})
is based on global properties of spacetime (existence of event horizon),
or thermal field theory (periodicity in the propagator) \cite{HarHaw,GibPer}.
In 1987 \cite{HuEdmonton} I proposed an alternative approach,
viewing it as a {\it local, kinematic effect} arising from the vacuum
being subjected to an {\it exponential scale transformation}. I stressed
that the salient features of inflation and black hole collapse
are dominated by the {\it infrared behavior} of quantum fields,
an effect O'Connor and I \cite{HuOC} called `dynamical' finite size effect.
Later, with Zhang, I \cite{cgea} used the
analogy with the Kadanoff-Migdal tranformation in critical phenomena \cite{KM}
to describe an inflationary universe,
and proposed to view  the late stage of inflation and black hole collapse
as approaches to the critical regimes \cite{jc}.

Here, I shall further develop this view by generalizing to {\it relativistic
exponential scaling} and, using the recently developed theory of {\it noise from
quantum fluctuations}, show how these effects viewed in one vacuum
(Rindler, Hartle-Hawking, or Gibbons-Hawking in the cases of uniformly
accelerated detector, black hole and de Sitter universe) can be understood
as resulting from the scaling of quantum noise in another (Minkowski).
Depicting Hawking-Unruh effect in the light of quantum and thermal fluctuations
was first proposed by Sciama \cite{Sciama}.
As a pedagogical illustration (mainly for audiences of field theory and
statistical mechanics, not general relativity), I present a simple derivation
of the de Sitter metric by invoking only the scale transformation concept
and  special-relativistic effects,
and show that Hawking radiation in these cases can be viewed as arising from
exponential-redshifting.
Though not new, our emphasis on the kinematic aspect over
the conventional global geometric  or thermal field aspects is intended
to clear the conceptual and technical pathways in order to introduce
non-equilibrium statistical field theory
\cite{HPZ,Ang,HM2,RHA,HMLA,HuBelgium,Banff,nfsg,HM3,fdrsc}.
In our view this theory based on the use of open system concepts \cite{qos} and
influence functional methods \cite{if} is more suitable for treating
conditions not easily amenable by traditional methods,  such as systems which
do not have an event horizon.
We have started this
investigation recently. Analysis of typical cases \cite{RHK,RHKM}
shows that in such systems radiance indeed is observed, albeit not in an exact
Planckian spectrum.
The ideas of exponential scaling \cite{HuEdmonton} and results on the infrared
behavior of quantum fields \cite{HuOC}
can be applied to the analysis of critical phenomena possibly occuring
in black holes, early universe and in the semiclassical to quantum gravity
transitions.

The stochastic approach we have adopted  bears in relation to black hole
thermodynamics \cite{Davies,Wald} the same way as stochastic and
non-equilibrium statistical mechanics bear to equilibrium thermodynamics.
A near-thermal Unruh radiation from  an almost-uniformly accelerated detector
and a black hole with its modified radiation can be viewed as the
linear-response regime, which can be treated by perturbation theory,
as distinguished from the fully non-equilibrium conditions of
dynamical collapse or arbitrary trajectories, where, we believe, the stochastic
theory of particles and fields will prove to be more useful than the
conventional methods.
(Logically stochastic mechanics connects the foundation, i.e.,
information and probability theory, to kinetic theory,
statistical mechanics and thermodynamics.
For an explanation of these conceptual points see \cite{dch,cddn}).
Our approach thus puts more emphasis on  statistical mechanics and
field theory than on geometry. This has both a technical and a conceptual
rationale. We find the geometric description of spacetime a very elegant,
but rather restrictive one, most effective in the large scale,
near-equilibruim state of matter and spacetime.
I have been of the opinion \cite{HuGRhydro} that
Einstein's theory of general relativity describes only the hydrodynamic
or thermodynamic regime of spacetime structure. If we want to probe into
the microscopic structure of spacetime and matter, we need to use quantum
field theory. (When combined with classical spacetime it defines
the semiclassical gravity theory \cite{BirDav} which
was instrumental to the discovery of the Hawking-Unruh effect.)
If we want to probe into the dynamical and collective
properties of spacetime and matter, we need to use non-equilibrium
statistical mechanics \cite{Spohn}. Thus we prefer to use the statistical and
stochastic quantum field theoretical approach to tackle the new generation
of problems like backreaction of particle creation in dynamical collapse,
statistical entropy and information loss puzzle in black holes, phase transition
in the early universe, and cross-over from semiclassical to quantum gravity.
We believe that the usual geometric depiction can be
recovered, and is only valid, in the equilibruim and long wavelength
(thermo- and hydro-dynamics) limits.

\section{Relativistic Exponential Scale Transformation}

In this section I will show how exponential scale transformation enters in
a central way in the class of spacetimes which admit Hawking-Unruh
radiation.   To highlight this as a kinematic effect, I'll use just special
relativity to derive the metric of such a spacetime.
There is nothing new in the result, which is well-known in general relativity.
But this alternative viewpoint and the derivation of an effect usually regarded
as belonging to the realm of general relativity could be of some interest for
audiences in statistical mechanics and field theory.
In this spirit, we will use the ideas of scaling from critical phenomena
in conjunction with a quantum theory of noise from stochastic field theory
to explore the deeper meaning of these effects and thus prepare the stage for
dealing with more complex problems.

\subsection{Nonrelativistic scaling}

Consider two systems S and $\tS$. S is stationary and  $\tS$ moves with
speed $\beta$ with respect to S.
Let the origins of these two systems coincide at $t= \tt =0$, and denote
the lengths measured in the two systems as $r, \tr$ respectively.
They are related by
\be
\tr = a(t) r
\te
where $a(t)$ is the scale factor relating the two systems. (One can evoke
a picture of polka dots on a balloon which is being inflated, such as in
Fig. 27.2 of \cite{MTW}, the distance between any neighboring dots
increases with time. Indeed one can interchangeably use $a(t)$ as
a measure of time.) Consider the special class of time dependence:
\be
a(t) = e^{Ht}
\te
If, say, at each unit time interval $\Delta t$ the length in $\tS$ doubles,
i. e., at $t=0, \tr=r; t=1, \tr= 2r; t=2, \tr=4r, ...$ Then $H \Delta t = ln 2$.
The exponential transformation is a special class, because
the rate of expansion $H \equiv \dot a/a$ is independent of time. (Compare to
a power law $a(t)= t^p$, $H= p/t$ depends on $t$. This seemingly minor
point actually makes a great difference in the relation of quantum and
thermal fluctuations in these two systems, as we shall see later.)
From this we get
\be
H\tr = Har = \dot a r = \beta,
\te
which is the relative velocity between the S and $\tS$ systems.
It is also the velocity the dots are receding from the origin.
Each dot on the expanding
balloon sees every other dot moving away from it isotropically,
the farther the distance the faster.

From this simple kinematics two related pictures might be evoked:\\
1) An isotropically expanding universe depicted by
the Friedmann-Robertson-Walker model. In particular, if $a(t) = e^{Ht}$,
it is the inflationary universe \cite{infcos}.\\
2) For the exponential expansion, the Kadanoff-Migdal (KM) transformation.

Recall that in critical phenomena \cite{KM} the KM transform is used in
conjunction with a block-spin transformation to render a problem defined on
a lattice with spacing $l$ to that of one defined on a larger lattice
(after $n$ iterations) with spacing $[a(\Delta t)]^n l$, and rescaled bond
strength. In the above example, $a(\Delta t) = e^{H \Delta t}= 2$.
Near the critical point where many systems manifest scaling behavior,
the transformed system preserves the long range characteristics of the
original system.
This is one way how one could use the ultraviolet behavior of the system
to analyze its infrared behavior (via the renormalization group equation,
usually constructed from the counterterms introduced for the removal
of ultraviolet divergence \cite{KM}).
I have used this analogy to explain the scale-invariant properties of
inflationary cosmology, and advocated that
one can understand the end state of inflation and black hole collapse
in terms of the approach to the critical regime, where the infrared behavior of
quantum fields in these spacetimes become dominant.

\subsection{Relativistic Scaling}

In the above, the relation $\tr = a(t) r$ assumes absolute time, i.e.,
S and $\tS$ use the same time. Thus the usual KM transformation is
non-relativistic scaling. This is of no surprise as there is no concern for
relativistic covariance in ordinary critical phenomena studies. 
Here, to respect special relativity, the two times should be related by a
Lorentz factor $\gamma$, i.e.,
\be
{{a(\tt)} \over {a(t)}} = \gamma = {1 \over {\sqrt{1 - \beta^2}}}
\te
To make precise the terminology, we shall call the following cases defined by
\\
Eq. (2.1) as nonrelativistic scale transformation (or scaling, for short),\\
Eq. (2.1) and (2.2) as exponential scale (or KM) transformation,\\
Eq. (2.1) and (2.4) as relativistic scaling, \\
Eq. (2.1) and (2.2) and (2.4) as relativistic exponential scaling, or
relativistic KM transformation.\\

Let us examine the meaning of Conditions (2.3) and (2.4).  Note that\\
i) When $\beta = 0$, or $\gamma =1$ i.e., $H=0,  (a$ a constant, $\tr = r$) or
   $\tr = 0$ (at the origin, $\tr=r$, $a(\tt)= a(t), t= \tt$. The two systems are
   identical.\\
ii) When $\tr \neq 0$ or $H \neq 0, \beta \neq 0$, then $a(\tt) = \gamma a(t)$.
    The frequency of radiation $ \tilde \nu $ measured in $\tS$ is redshifted
    from $\nu$ measured in S by a factor $\gamma$. Thus $\gamma$ is the
    redshift factor. For exponential scaling, $H=constant$,
    we have exponential red-shifting.\\
iii) When $\beta \rightarrow 1$ or at $\tr = H^{-1}, \gamma \rightarrow 
    \infty$.    There exists an event horizon of infinite redshift.\\

Let us now see how the redshift factor can be attributed to a non-flat metric.
The easiest way is to recall how gravitational red-shift
can be viewed as a special relativistic effect.  It is determined by the
$g_{00}$ component of the metric.
More precisely,
\be
\tilde g_{00} = 1 - \beta^2,~~~ or,~~~ \sqrt{\tilde g_{00}} = \gamma^{-1}.
\te
In the example of a (2D) Schwarzschild metric for a massive (M) object,
\be
ds^2 = (1- {{2M}\over{r}}) dt^2 - {{dr^2}\over{(1- {{2M}\over{r}})}},
\te
we can see the following analogous cases:\\
i) At $r = \infty$, $g_{00} = 1, \gamma =1 $,
the metric is asymptotically flat.\\
ii) At $r> 2M, ~~1 > g_{00} > 0, ~~\gamma > 1$.\\
iii) At $r=2M, g_{00}= 0, \gamma \rightarrow \infty$.
This infinite-redshift surface is, of course, what defines the black hole
horizon.

Hence by analogy, we see that for the coordinates $(\tt,\tr)$ in the
$\tS$ frame, $\tilde g_{00} = \gamma ^{-2} = 1 - (H \tr)^2$. The metric
embodying an exponential scale transformation is thus of the form:
\be
ds^2 = [1- (H\tr)^2] d \tt^2 - {{d \tr^2 }\over {[1- (H\tr)^2 ]}}
\te
This is the form of a de Sitter metric in the so-called static coordinate.
Note that an event horizon exists at $\tr = H^{-1}$
for observers at $\tr = 0$. The exponential red-shifting factor from
relativistic scaling is what we prefer to focus on (rather than
the geometric structure) in the statistical field theory
approach to thermal effects and their generalizations.

\section{Thermal Particle Creation from Exponential Red-shifting}


Particle creation from a black hole (Hawking effect) or observed by a uniformly
accelerated observer (Unruh effect) have been treated in many ways by
methods in quantum field theory in curved spacetimes \cite{BirDav}.
Of the many existing derivations and interpretations, I want to highlight what I see
as the central aspect of this problem, i.e.,
the vacuum state where  thermal radiance is observed (e.g., the Rindler
vacuum of a constantly accelerated detector in Unruh effect) is related to
the inertial vacuum state (Minkowski vacuum) by an exponential scale
transformation.
This transformation is of the same form and nature as
that between the $S, \tS$ systems in the simple example given above.
In fact, the radiation observed by an observer in the $\tS$ system is also of
a thermal nature. It was first deduced by Gibbons and Hawking \cite{GibHaw} by
analogy with the Hawking radiation in a Schwarzschild metric. In all
three cases, i.e., the Rindler, the Schwarzschild and the de Sitter
metrics, it is the exponential red-shifting which is responsible for
the thermal nature of radiation. We make the fine distinction that it is
the exponential redshifting and not the existence of an event horizon
which is the necessary condition. The above situations encompass both the cases
where there is an event horizon but no (thermal) particle creation
(like an extreme Reisner-Nordstr\"om metric), and cases where the horizon
is not globally defined (like the case of finite-time acceleration).
The former case was brought up in a discussion between me and Unruh during
his visit to Maryland in 1988 but never pursued.
The latter case is treated by Raval, Koks and myself \cite{RHK} recently,
using the statistical field theory methods.

Consider the case of a uniformly accelerated detector. If we write the Minkowski
metric in the null coordinates (U, V)
\be
ds^2 = dt^2 - dx^2 = dU dV
\te
where
\be
U = t - x, ~~ V = t + x,
\te
and perform a conformal transformation to the Rindler coordinates
$(\xi, \eta)$,
\be
ds^2 = e^{2a \xi} (d\eta^2 - d \xi^2)
\te
with the associated null coordinates $(u, v)$
\be
u =   \eta - \xi, ~~~ v =   \eta + \xi,
\te
the two sets of null coordinates are then related by
\be
U = - \frac{1}{a} e^ {-au}, ~~~ V = \frac{1}{a} e^{av}.
\te
Particle detector moving at constant acceleration ($ \xi = const $) has a
trajectory $ x^2 - t^2 = \alpha^2  $, where
$ \alpha^{-1}  =a e^{-a\xi} $ is the proper acceleration. The detector's
proper time is $ \tau  = e^{a \xi} \eta$. (See discussion in, e.g.,  Sec. 4.5
of \cite{BirDav})

The Schwarzschild metric (2.6) depicing an eternal (2D) black hole can be
written in terms of the Regge-Wheeler coordinates
\be
r^* = r + 2M ln | (\frac{r}{2M}) - 1|
\te
as
\be
ds^2 = (1- \frac{2M}{r}) (dt^2 -dr^{*2}) = (1- \frac{2M}{r}) du dv,
\te
where $(u, v) = t - r^*, v= t + r^* $ are the null Schwarzschild
coordinates. In analogy with the uniformly accelerated observer case,
one can introduce a set of Kruskal coordinates, $(\bar t,  \bar r^*)$,
and write the metric as (see, e.g., Sec. 3.1 of \cite{BirDav})
\be
ds^2 = \frac{2M}{r} e^{- \frac{r}{2M}} (d \bar t^2 - d \bar r^{*2})
     = \frac{2M}{r} e^{- \frac{r}{2M}} (dU dV)
\te
The null Schwarzschild and  Kruskal coordinates are related by
\be
U = -4M e ^{-\frac{u}{4M}}, ~~~ V = 4M e ^{\frac{v}{4M}}
\te

Note the pairwise correspondence between the Minkowski / Kruskal $(U, V)$
and the Rindler / Schwarzschild $(u, v)$ coordinates in the accelerated
detector and the black hole cases. Note again the exponential relation
between these two sets of coordinates.

Let me briefly describe the relation between  Killing vectors, normal modes,
vacuum states, Bogolubov transformation and particle creation in a curved
spacetime. (The reader is referred to e.g., \cite{BirDav} for details).
The existence of a Killing vector in a spacetime (e.g., $\partial _t$,
or equivalently, $\partial_U, \partial_V$ in Minkowski space) allows for a
normal mode decomposition. The amplitudes of the normal modes when second
quantized define the creation and annihilation operators ($A, A^+$),
and the number operator $n = A^+ A$ which make up the Fock space
with respect to this decomposition. The vacuum $|0>_t$ being the no particle
state is defined by taking modes to be positive frequency with respect
to the Killing vector $\partial_t$.
The amplitude functions or the annihilation and creation
operators ($a_j, a^+_j$) of another set of modes decomposed
with respect to another Killing vector (e.g. $\partial_u$) is related to
the original one by a Bogolubov transformation:
\be
a_j = \Sigma_i (\alpha_{ij} A_i + \beta_{ij}^* A_i^+)
\te
where $\alpha, \beta$ as the Bogolubov coefficients of the $kth$ mode.

In the case of a collapsing mass, an incoming wave from past infinity
${\cal I}^-$ in the form $e^{-i\omega v}$ (the Killing vectors
$\partial_u, \partial_v $ define the Schwarzschild or the Boulware vacuum)
falling towards the mass would be subjected to a blue-shift. Having passed
through the collapsing mass, the outgoing wave while climbing out of the
severe and increasing gravitational potential is subjected to an exponential
redshifting which far exceeds the blue-shift (there is also a small
spin-dependent contribution from the effective potential), i.e.,
\be
e^{-i\omega v} \rightarrow e^{+i\omega U} 
= e^{+i\omega ( -4M e ^{-\frac{u}{4M}})}
\te
It is seen that the exponential arises from the defining relation of $(u, v)$
and $(U, V)$, (the Killing vectors $ \partial_U, \partial_V $ are used in
the case of an eternal black hole -- black hole in equilibrium with its Hawking
radiation -- to define the Hartle-Hawking vacuum).
The so-called Unruh vacuum (defined with an `in' state with respect to
$\partial_v$ and `out' state with respect to $\partial_U$) is
most suitable for the description of the actual black hole collapse and
radiation emittance situation.

For the above set-up,
the coefficient $\beta$ connecting a positive frequency incoming component
$e^{-i\omega v}$ and a negative frequency outgoing component
$e^{+i\omega ( -4M e ^{-u/4M})}$ has the special form:
\be
|\beta_k /\alpha_k|^2 = e^{-8\pi M \omega}~~or ~~ e^{-2\pi \omega /a},
\te
where the second term is for the accelerated observer. This
leads to a Planckian spectrum
\be
<n_k> = ( e^{8\pi M \omega} - 1)^{-1}~~ or ~~ ( e^{2\pi \omega /a} -1) ^{-1},
\te
which gives respectively the Hawking and Unruh temperatures $T_H, T_U$
\be
k_B T_H = \frac{1}{8 \pi M}~~, ~~ k_B T_U = \frac{a}{2 \pi},
\te
where $k_B$ is the Boltzmann constant.
Note that this form is readily identifiable from the exponential form of the
relations (3.5) (3.9) between $(u, v)$ and $(U, V)$ defining the in and out states.
The out wave being exponentially redshifted is responsible for the thermal
nature of the Hawking-Unruh radiation.

Let us now return to the observers in S and $\tS$. The two vacuum states
$ |0>_S, |0>_{\tS} $ defined with respect to the two Killing vectors 
$\partial_t, \partial_{\tt}$ bear  the same relation as the
Schwarzschild versus the Kruskal vacuum. By analogy, an observer in $\tS$
will therefore see a thermal radiation with temperature given by
\be
k_B T_{dS} = \frac{H}{2 \pi}.
\te
(This  was first derived by Gibbons and Hawking \cite{GibHaw}.)
We use the de Sitter space  for illustration because it
manifests the kinematic effect most directly. But for all three cases, we
can say that thermal radiance detected by one observer (in $\tS$ )
arises from the relativistic exponential scaling of vacuum fluctuations of
the other (in S).

To complete our thesis, it remains to show how vacuum
fluctuations can be understood as quantum noise. For this we need to
introduce some basic notions and techniques of statistical field theory.
We will try to illustrate the basic ideas with miminal technical detail .


\section{Quantum Noise under Exponential Scaling Manifests as Thermal Radiation}

Noise from vacuum fluctuations \cite{HMLA}
can best be defined in terms of the stochastic
theory of quantum fields using the quantum open system concept and the influence
functional formalism. A general  review of this method is given in \cite{Banff}.
Details of the following summary are contained in \cite{HM2}.

We consider an Unruh-DeWitt detector undergoing constant acceleration $a$
with trajectory (in 2D)
\be
x(\tau) = \frac{1}{a} cosh a \tau, ~~ s(\tau) = \frac{1}{a} sinh a \tau
\te
where $(x, s)$ are its internal coordinates and $\tau$ its proper time.
For a 2D scalar quantum field $\Phi (x_0, \eta)$ in flat space, one can
decompose it in normal modes, each describable by a Lagrangian
\be
L (s) = \ha \sum_\sigma  \sum_k [ (\dot q_k^\sigma)^2 - k^2 (q_k^\sigma)^2]
 \te
where  $q_k^{\sigma}$ is the amplitude functions for the kth mode
(here $\sigma = \pm$ denotes the $sin$ and $cos$ standing wave components).
The detector-field interaction is described by an interaction Lagrangian density
\be
L_{int}(x) = - \epsilon r \Phi (x) \delta (x(\tau)).
\te
The influence of the quantum field on the detector is expressed in terms of an
influence kernel which has the form
\be
\zeta (s(\tau), s(\tau')) = \nu (s,s') + i \mu (s,s')
\te
where $\nu, \mu$ are the noise and dissipation kernels.
For an uniformly accelerated observer it is found by Anglin, Hu and Matacz
\cite{Ang,HM2} to be,
\be
\zeta (\tau, \tau') = \int_0^\infty  dk I(k)
[ coth (\pi k / a) cos k (\tau -\tau') - i sin k (\tau -\tau')],
\te
where $I(k)$ is the spectral density function.
By comparison with the Brownian motion model \cite{if},
it is seen to have the same form as a particle in a thermal bath with
temperature given by (3.14), the expected Unruh temperature.
Similarly, for a 2D black hole, we see that a detector at $\bar r^* =constant$
is influenced by the quantum field in the same way (same form of the influence
functional) as an inertial detector in flat 2D spacetime;
while a detector at $r^* = constant$ has the same influence
functional as an uniformly-accelerated detector in flat 2D spacetime.
Using this analogy, it is easy to see that the influence kernel of a
scalar field on a detector at $r^* = constant$
(which defines the Hartle-Hawking vacuum) has the form
\be
\zeta (t, t') = \int_0^\infty dk I(k) [ coth (4\pi M k) cos k (t -t')
- i sin k (t -t')].
\te
By comparison with a Brownian oscillator in a thermal bath, it is easily
seen that the detector will see a thermal radiation with the
Hawking temperature (3.14).

This derivation of the Hawking-Unruh radiation relies mainly on the statistical
field rather than geometric ideas. In this approach, it is seen
that the quantum noise  in one set of  vacuua (the Minkowski,  Kruskal or the
S observer) becomes thermal radiation in  another set of vacuua
(the Rindler, Schwarzschild or the $\tS$ observer).
The latter is related to the former by a relativistic exponential
scale transformation, the exponential red-shifting being
responsible for the exact thermal but coherent characteristics of the radiance.

One may think that the kinematic and statistical approach we have presented
here is just another description equivalent to the traditional geometric
approach. This is true for exactly thermal (equilibrium) conditions,
corresponding to the special case of uniform acceleration.
But we think the statistical field theory approach has an advantage over
the geometric approach in treating cases which deviate from these conditions,
e.g., for detectors which undergo acceleration for only a finite
interval of time, or approach uniform acceleration asymptotically,
as well as dynamical collapsing mass, and
cosmological models with near-exponential expansion.
In these cases there does not exist an event horizon, so the traditional
arguments depending on such a condition
(such as the periodicity in the thermal propagator)
would not be available. However, one can still use concepts such as
near-exponential scale transformation and perturbation techniques
in field theory to treat these cases.
In two recent papers, Raval, Koks, Matacz and I  \cite{RHK,RHKM}
have shown that in such systems radiance indeed is observed,
albeit not in a precise Planckian spectrum.
The deviation therefrom is determined by a parameter which
measures the departure from  uniform acceleration or from
exact exponential expansion.
These results are expected to be useful for investigating the non-equilibrium
black hole thermodynamics and the linear-response regime of quantum backreaction
problems.\\

\noindent {\bf Acknowledgement}
I wish to thank the organizers and hosts of this conference, Profs. Gui and
Su, for invitation to participate, and Prof. Yavuz Nutku of Tubutak, Turkey,
for his warm hospitality during my visit to Istanbul in July, 1995,
supported by a NATO grant, where this talk was prepared. I am also grateful
to Dr. Jonathan Simon for suggestions on improving the clarity of presentation
in this write-up.
Research is supported in part by the  U. S. National Science Foundation
under grant PHY94-21849.


\newpage

\begin{thebibliography}{999}

\bibitem {Haw75}
S.W. Hawking, Commun. Math. Phys. {\bf 43}, 199 (1975).

\bibitem {Unr76}
W.G. Unruh, Phys. Rev. {\bf D 14}, 870 (1976).
B. S. DeWitt, Phys. Rep. 19C, 297 (1975).
P. C. W. Davies, J. Phys. {\bf A}: Gen. Phys. {\bf 8}, 609 (1975).
S. A. Fulling, Phys. Rev. {\bf D 7}, 2850 (1973) .

\bibitem {FulDav}
S.A. Fulling, P.C.W. Davies, Proc.\ R. Soc.\ Lond.\ {\bf A 348}, 393 (1976).

\bibitem {HawRad75}
W. Israel, Phys. Lett. 57A, 107 (1975);
L. Parker, Phys. Rev. D12, 1519 (1975);
R. M. Wald, Comm. Math. Phys. 45, 9 (1975)

\bibitem {GibHaw}
G. Gibbons and S. W. Hawking, Phys. Rev. D15, 1752 (1977)

\bibitem {Par76}
L. Parker, Nature 261, 20 (1976)

\bibitem {Lap78}
A. Lapedes, J. Math. Phys. 19, 2289 (1978)


\bibitem {HarHaw}
J. B. Hartle and S. W. Hawking, Phys. Rev. D13, 2188 (1976)

\bibitem {GibPer}
G. Gibbons and M. J. Perry, Proc. Roy. Soc. Lon. A358, 467 (1978)

%

\bibitem {HuEdmonton}
B. L. Hu, in {\it Proceedings of the CAP-NSERC Summer Institute
in Theoretical Physics, Vol 2}, Edmonton, Canada, July 1987, eds F. C. Khanna,
G. Kunstatter and H. Umezawa (World Scientific, Singapore, 1988)

\bibitem{HuOC}
B. L. Hu and D. J. O'Connor, Phys. Rev. {\bf D36}, 1701 (1987).
Phys. Rev. Lett. 56, 1613 (1986)

\bibitem {cgea}
B. L. Hu and Y. Zhang, ``Coarse-Graining, Scaling, and Inflation"
Univ. Maryland Preprint 90-186 (1990);
B. L. Hu, in {\it Relativity and Gravitation: Classical
and Quantum} Proc. SILARG VII, Cocoyoc, Mexico 1990.
eds. J. C. D' Olivo et al (World Scientific, Singapore 1991).

\bibitem {KM}
S. K. Ma, {\it Modern Theory of Critical Phenomena} (Benjamin, London, 1976).
D. J. Amit, {\it Field Theory, the Renormalization Group and Critical Phenomena}
2nd ed. (World Scientific, Singapore, 1982)

\bibitem {jc}
B. L. Hu, ``Nonequilibrium Quantum Fields in Cosmology: Comments on
Selected Current Topics" in  {\it Second Paris Cosmology Colloquium}
Observatorie de Paris, June 2-4, 1994 eds H. J. de Vega and N. Sanchez
(World Scientific, Singapore, 1995) gr-qc/9409053

\bibitem {Sciama}  D. W. Sciama, ``Thermal and Quantum Fluctuations in
Special and General Relativity: an Einstein Synthesis'' in {\it Centenario
di Einstein} (Editrici Giunti Barbera Universitaria) (1979); P. Candelas and
D. W. Sciama, Phys. Rev. Lett. {\bf 38}, 1372 (1977); D. W. Sciama, P.
Candelas and D. Deutsch, Adv. Phys. {\bf 30}, 327 (1981).

\bibitem {HPZ}
B. L. Hu, J. P. Paz and Y. Zhang, Phys. Rev. {\bf D45}, 2843 (1992);
{\bf D47}, 1576 (1993)

\bibitem {Ang}
J. R. Anglin, Phys. Rev. {\bf D 47}, 4525 (1993) .

\bibitem {HM2}
B. L. Hu and A. Matacz, Phys. Rev. {\bf D 49}, 6612 (1994)

\bibitem {RHA}
Alpan Raval, B. L. Hu and J. R. Anglin,
``Stochastic Theory of Accelerated Detectors in a Quantum Field" (1995)
gr-qc/9510002

\bibitem{HMLA}  B. L. Hu and A. Matacz, ``Quantum Noise in Gravitation and
Cosmology'' in Proc. International Workshop on {\it Fluctuations and Order:
A New Synthesis}, Los Alamos, Sept. 1993, Proceedings edited by Marko
Millonas (Springer-Verlag, Berlin, 1996). University of Maryland Preprint
umdpp94-44. astro-ph/9312012

\bibitem{HuBelgium}
B. L. Hu, J. P. Paz and Y. Zhang, ``Quantum Origin of Noise and Fluctuation
in Cosmology'' in {\it The Origin of Structure in the Universe}
Conference at Chateau du Pont d'Oye, Belgium, April, 1992,
ed. E. Gunzig and P. Nardone (NATO ASI Series) (Plenum Press, New York,
1993) p. 227.  gr-qc/9512049

\bibitem {Banff}
B. L. Hu, ``Quantum Statistical Fields in Gravitation and
Cosmology'' in {\it Proc. Third International Workshop on Thermal Field
Theory and Applications}, eds. R. Kobes and G. Kunstatter (World Scientific,
Singapore, 1994) gr-qc/9403061

\bibitem {nfsg}
E. Calzetta and B. L. Hu, Phys. Rev. {\bf D 49 }, 6636 (1994)

\bibitem {HM3}
B. L. Hu and A. Matacz, Phys. Rev. {\bf D 51}, 1577 (1995)

\bibitem {fdrsc}
B. L. Hu and S. Sinha,
Phys. Rev. {\bf D 51}, 1587 (1995).

\bibitem{qos}  See, e.g., E. B. Davies, {\it The Quantum Theory of Open
Systems} (Academic Press, London, 1976); K. Lindenberg and B. J. West,
{\it The Nonequilibrium Statistical Mechanics of Open and Closed Systems}
(VCH Press, New York, 1990); U. Weiss, {\it Quantum Dissipative Systems}
(World Scientific, Singapore, 1993)

\bibitem{if}
R. Feynman and F. Vernon, Ann. Phys. (NY) {\bf 24}, 118 (1963).
R. Feynman and A. Hibbs, {\it Quantum Mechanics and Path Integrals},
(McGraw - Hill, New York, 1965).
A. O. Caldeira and A. J. Leggett, Physica {\bf 121A}, 587 (1983).
H. Grabert, P. Schramm and G. L. Ingold, Phys. Rep. {\bf 168}, 115 (1988).
B. L. Hu, J. P. Paz and Y. Zhang, Phys. Rev. {\bf D45}, 2843 (1992);
{\bf D47}, 1576 (1993)

\bibitem {RHK}
A. Raval, B. L. Hu and D. Koks,
``Stochastic Theory of Near-Thermal Radiation in Accelerated Detectors, Mirrors,
and Black Holes" (1996)

\bibitem{RHKM}
A. Raval, B. L. Hu, D. Koks and A. Matacz
``Stochastic Theory of Near-Thermal Particle Creation in Expanding Universes"
(1996)

\bibitem {Davies}
P. C. W. Davies, Rep. Prog. Phys. 41, 1313 (1978).

\bibitem {Wald}
R. M. Wald, {\it Quantum Field Theory in Curved Spacetime and Black Hole
Thermodynamics} (Univ. of Chicago, Chicago, 1994)

\bibitem{dch}  E. Calzetta and B. L. Hu, ``Decoherence of Correlation
Histories'' in {\it Directions in General Relativity, Vol II: Brill
Festschrift}, eds B. L. Hu and T. A. Jacobson (Cambridge University Press,
Cambridge, 1993) gr-qc/9302013

\bibitem {cddn} E. Calzetta and B. L. Hu, ``Correlations, Decoherence,
Disspation and Noise in Quantum Field Theory'', in {\it Heat Kernel
Techniques and Quantum Gravity}, ed. S. A. Fulling (Texas A\& M Press, College
Station 1995). hep-th/9501040

\bibitem {HuGRhydro}
B. L. Hu, ``General Relativity as Hydrodynamics"  seminars given at
the University of Maryland,  Spring, 1994 (unpublished). Talk at the
Second Sakharov Conference, Moscow, May, 1996.

\bibitem {BirDav}
N. D. Birrell and  P. C. W. Davies, {\it Quantum Fields in Curved Space}
(Cambridge University Press, Cambridge, 1982).

\bibitem {Spohn}
H. Spohn, {\it Large Scale Dynamics of Interacting Particles}
(Springer-Verlag, Berlin, 1991)

\bibitem {MTW}
C. W. Misner, K. S. Thorne and J. A. Wheeler, {\it Gravitation}
(Freeman, San Francisco, 1973).

\bibitem {infcos}
 A. H. Guth, Phys. Rev. D 23, 347 (1981).
 A. Albrecht and P. J. Steinhardt, Phys. Rev. Lett. 48, 1220 (1982).
 A. D. Linde, Phys. Lett. 114B, 431 (1982). Phys. Lett. 162B, 281 (1985).

\end{thebibliography}
\end{document}